\documentclass[reprint,showpacs,showkeys,prb,aps,eqsecnum]{revtex4-1}

\usepackage{graphicx}
\usepackage{epstopdf}
\usepackage{bm}
\usepackage{amsmath}
\usepackage{subfigure}
\usepackage{mathrsfs}
\renewcommand{\vec}{\bm}

\begin{document}

\title{Spin-Wave Resonance Model of Surface Pinning \\
in Ferromagnetic Semiconductor (Ga,Mn)As Thin Films}

\author{H.~Puszkarski}
	\email{Corresponding author, email: henpusz@amu.edu.pl}
	\affiliation{Surface Physics Division, Faculty of Physics, Adam Mickiewicz University \\
					ul.~Umultowska 85, 61-614 Pozna\'n, Poland}

\author{P.~Tomczak} 
        \email{email: ptomczak@amu.edu.pl}
	\affiliation{Quantum Physics Division, Faculty of Physics, Adam Mickiewicz University \\
ul. Umultowska 85, 61-614 Pozna\'n, Poland}

\date{\today}

\begin{abstract}
The source of spin-wave resonance (SWR) in thin films of the ferromagnetic semiconductor (Ga,Mn)As is still under debate: 
does SWR stem from the \emph{surface} anisotropy (in which case the surface inhomogeneity (SI) model would apply), 
or does it originate in the \emph{bulk} inhomogeneity of the magnetic structure of the sample (and thus requires the use
of the volume inhomogeneity (VI) model)? This paper outlines the ground on which the controversy arose and shows why in different 
conditions a resonance sample may meet the assumptions of either the SI or the VI model. 
\end{abstract}


\maketitle


\section{Introduction}\label{sec_intro}

Dilute ferromagnetic semiconductors are a class of very promising materials of the future. \cite{Furdyna1988,ohno_sci,dietl_nature,dietl_prb,jungwirth,dietl2010} Gallium manganese arsenide (Ga,Mn)As, created on the basis of the semiconductor gallium arsenide by the addition of a small percentage of manganese as a magnetic dopant, is one of the most intensively studied compounds in this class.\cite{ohno1996,konig,sawicki2003,sawicki2005,titova,liu06,bouzerar,werpach,nemec} The~free motion of positive charge carriers (holes) throughout the crystal results in the ferromagnetic order of the manganese ions. The basic magnetic characteristics of the material depend on the amount of the dopant ions and the spatial distribution of the concentration of the charge carriers (holes) transmitting magnetic information between the Mn ions. A particularly interesting situation occurs in thin films, in which magnetic characteristics (magnetic anisotropy, magnetization, exchange length and  stiffness constant, damping constant, etc.) are in general nonuniform along the growth direction perpendicular to the film surface. The character of this nonuniformity reflects the distribution profile of the charge carrier concentration in the film. 

The spatial magnetic profiles in thin films can be determined by means of ferromagnetic resonance, which reveals its fine structure in a \emph{multi-peak} resonance spectrum in thin-film systems; this type of ferromagnetic resonance is referred to as \emph{spin-wave resonance} (SWR), as each peak in the resonance spectrum corresponds to the excitation of a specific spin wave. On the other hand, the spectrum of allowed spin-wave excitations is determined by the shape of the \emph{magnon potential} of the system. Since the position of each SWR peak corresponds to a spin-wave energy level resulting from the prevailing magnon potential, an experimental SWR spectrum can be turned into the corresponding profile of the magnon potential by an appropriate calculation procedure. Thus, providing information on the spatial distribution of the basic magnetic characteristics, including the charge carrier concentration in the film, resonance measurements are of vital importance for the elucidation of the origins of ferromagnetism in the material under investigation.

Spin-wave resonance in thin films has been studied particularly intensively in gallium manganese arsenide in the past decade. \cite{sasaki2002,sasaki2003,liu2003,rappoport,liu2005,liu2006,liu2007,zhou2007,zhou_aip_conf,
zhou2009,goennenwein,bihler2009,dreher}
Especially rich resonance spectra were obtained in studies with a variable configuration of the static field with respect to the film surface. The field was rotated both perpendicularly to the film surface (which corresponds to variable polar angle $\theta_H$  between the direction of the external field and the surface normal) and in the plane of the film (variable azimuth angle $\phi_H$  between the external field and a reference direction in the film plane). The results of these measurements clearly indicate that the evolution of the SWR spectrum with the field configuration is correlated with that of the spatial distribution of the spontaneous magnetization and the anisotropy; thus, configuration and space dependence of the magnon potential should be assumed as well. 

In the present paper we shall only analyze SWR measurement data concerning the out-of-plane rotation of the magnetic field, mainly because of the controversy that arose in the interpretation of these results over an issue which therefore requires elucidation (in a separate paper we intend to analyze measurement data obtained in SWR studies with in-plane rotation of the magnetic field as well). If researchers tend to agree on the interpretation of SWR spectra in two extreme configurations -- the perpendicular and parallel configurations, corresponding to $\theta_H = 0$ and $\theta_H = 90^{\circ}$, respectively -- the interpretation of results obtained in intermediate configurations is under debate. Almost as a rule, a particular configuration of the external field tends to occur in this range at a critical angle $\theta_H^c$, for which the multi-peak SWR spectrum collapses to a single-peak FMR spectrum. There are two schools of thought regarding the interpretation of the occurrence of this critical angle.  These two prevalent opinions agree on the physical state of the thin film in the critical configuration, but differ in the interpretation of the configuration-related processes that accompany the rotation. Both schools agree that in the critical configuration the thin film (its magnon potential, to be precise) is \emph{magnetically} homogeneous, and the boundary conditions (specifically, the surface spin pinning) correspond to the natural conditions, only resulting from the reduced neighborhood of the surface spins (a precise definition of the natural pinning conditions is provided in the next Section). The difference of opinion concerns the configuration evolution leading to the above-described ``naturally homogeneous'' magnetic state. One school \cite{liu2007}
uses the surface inhomogeneity (SI) model and assumes that rotation of the magnetic field does not modify the profile of the bulk magnon potential, which remains homogeneous across the film; only the surface pinning conditions change, diverging from the natural conditions as the angle grows above or decreases below the critical configuration (with the surface pinning decreasing or increasing). In contrast, the other school, \cite{dreher}
using the volume inhomogeneity (VI) model, claims that it is the bulk profile of the magnon potential that changes with the field configuration: remaining \emph{linear}, but inclined at different angles with respect to the surface of the film, the magnon potential increases or decreases inside the film as the configuration diverges from the critical angle, while the natural conditions prevail invariably on the surface. In this paper we opt for the interpretation based on the SI model and propose a theoretical model of the configuration evolution of the surface spin pinning in agreement with the experimental data. Our interpretation leads to some physical conclusions, which provide new insights into the surface properties of ferromagnetic semiconductor (Ga,Mn)As thin films. 

\section{The goal of the study and the concept of SWR surface pinning parameter}\label{goal}

Our discussion of the state of the art of the research in the critical angle effect in SWR in ferromagnetic semiconductor (Ga,Mn)As thin films will rely on the representative study performed by Liu~\textit{et~al.}, reported in Ref.~\onlinecite{liu2007}, presenting SWR spectra measured for intermediate angles $\theta_H$  between the external field and the surface normal. Characteristically, in the \emph{out-of-plane configuration}, with the field vector rotated in a plane perpendicular to the surface, the SWR spectrum, consisting of multiple peaks in the perpendicular ($\theta_H =0$) and parallel ($\theta_H =90^{\circ}$) configurations, is found to collapse to a single-peak FMR spectrum in an intermediate configuration corresponding to a critical angle $\theta_H^c$ ($19^{\circ}$ in the studied sample).

The critical angle effect in SWR has been known for years, but that observed in (Ga,Mn)As samples is very unusual. The peculiarity is that the critical angle~$\theta_H^c$  coincides with the border between two configuration domains in which the SWR spectrum fulfills the assumptions of different models: the surface inhomogeneity model \cite{kittel}
for~$\theta_H >\theta_H^c$ (in which range the spacing between the resonance peaks is proportional to~$n^2$, where~$n$ is the spin-wave mode number), and the volume inhomogeneity model, \cite{portis}
which applies for~$\theta_H <\theta_H^c$ (where the spacing between the resonance modes is proportional to~$n$). A question arises: what mechanism underlies the occurrence of the inhomogeneity, if surface inhomogeneity prevails for~$\theta_H >\theta_H^c$, and volume inhomogeneity for~$\theta_H <\theta_H^c$? And what particular surface mechanism leads to the occurrence of the critical angle~$\theta_H^c$  at which these two types of inhomogeneity fail to be ``seen'' in the resonance? 

It should be noted that the SWR studies of (Ga,Mn)As conducted so far tended to focus on \emph{volume} characteristics only, such as the uniaxial anisotropy or the exchange constant of the studied material. The aim of this paper is to use SWR for getting a better insight into the ferromagnetism of dilute semiconductors in terms of their surface characteristics, the current knowledge of which is scarce. For this reason, in the analysis presented in this paper, we refer to our earlier quantum theory of SWR,
\cite{ferchmin1962,HP1970,yu,wigen+HP,HP1977,crack,HP1979,HP1981,ferchmin+HP,HP+MKasp} 
in which we have introduced the concept of \emph{surface spin pinning parameter}, a quantity that measures
 the degree of pinning of the surface spins and reveals explicitly different surface magnetic anisotropies 
present in thin films. 

\begin{figure}[h]
\centering
      \subfigure[\,Natural freedom]{\includegraphics [width=4 cm]{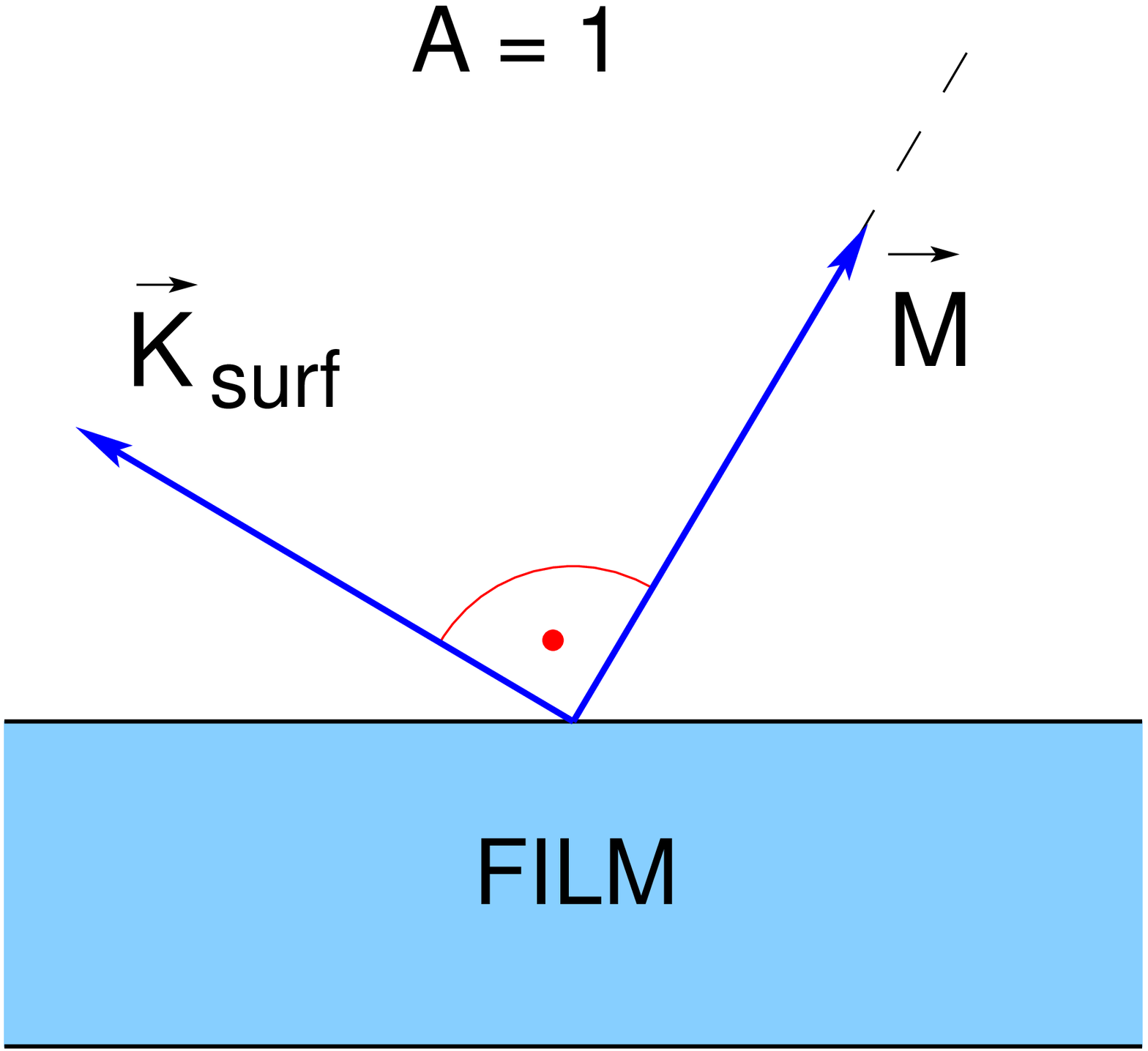}}
      \hspace{1cm}
      \subfigure[\,Pinned surface spins]{\includegraphics [width=4 cm]{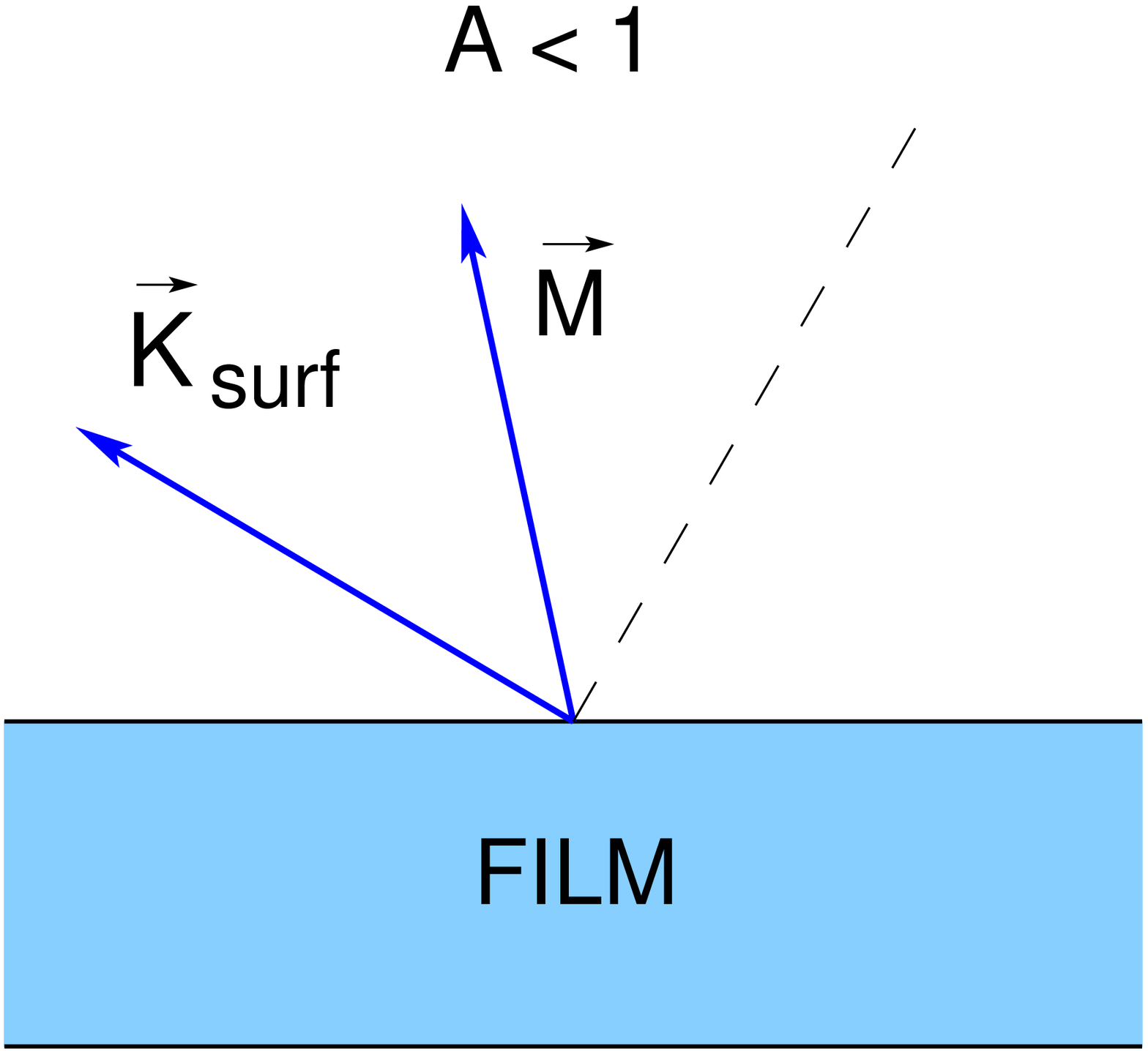}}
      \hspace{1cm}
      \subfigure[\,Unpinned surface spins]{\includegraphics [width=4 cm]{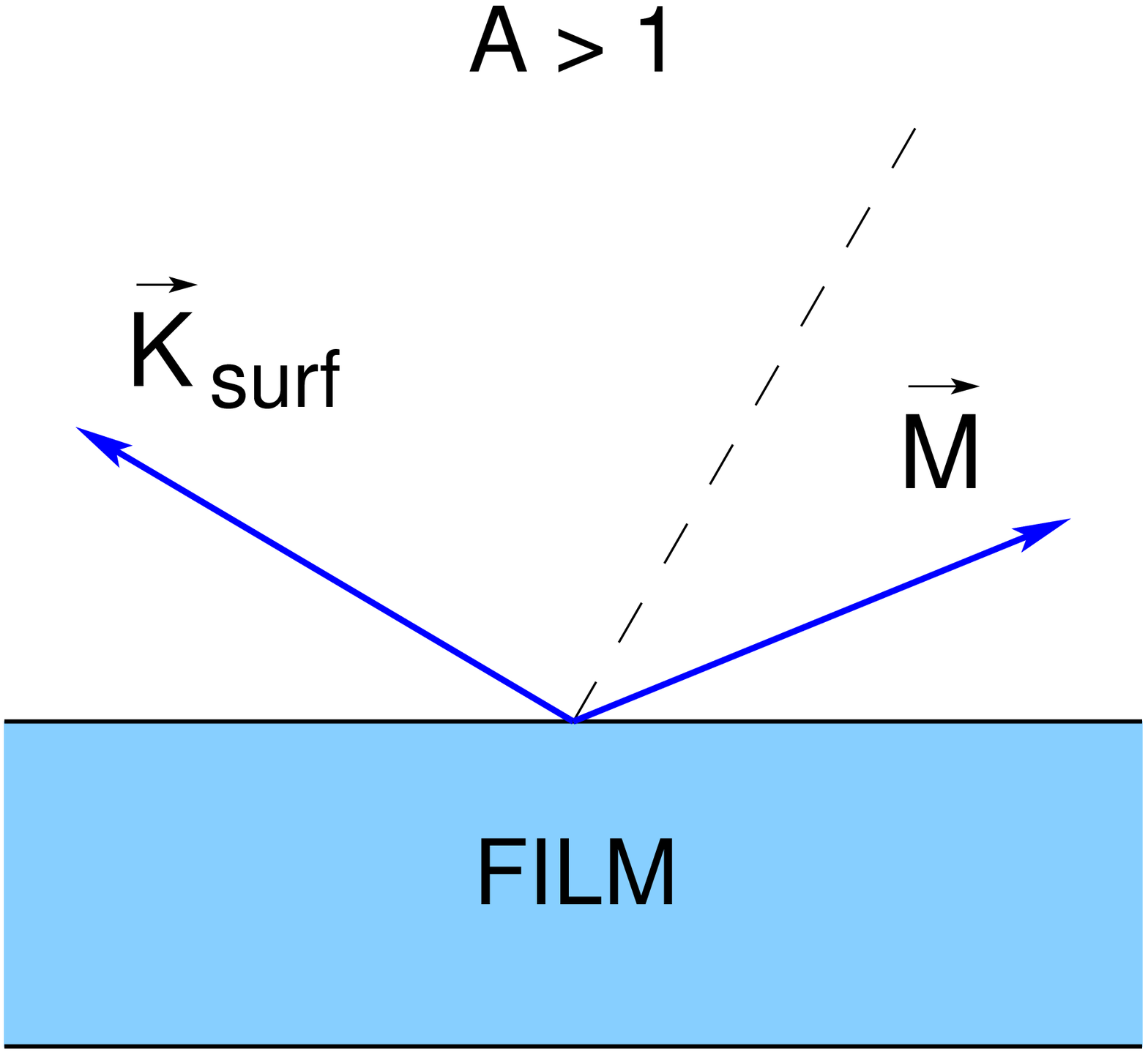}}
      \caption{Schematic representation of three surface spin pinning regimes which prevail in a thin film depending on the configuration of its magnetization $\vec{M}$ with respect to the effective surface anisotropy field~$\vec{K}_{surf}$ (see~(\ref{def_parameter})). 
When aligned as in~(a), the surface spins do not feel the anisotropy field and $A=1$, which corresponds to their natural freedom. In the configurations~(b) and (c) the surface spins are pinned ($A<1$) and unpinned ($A>1$), respectively, due to the anisotropy field.}
	\label{pinning_regimes}
\end{figure}

The concept of surface pinning is related to the descriptiof he energy status of surface spins, 
specifically to the degree of freedom of their precession. In a very simplified 
image introduced in Refs.~\cite{HP1970,HP1979},
besides the effective magnetic field present throughout the sample, an additional magnetic field $\vec{K}_{surf}$, referred to as the effective surface anisotropy field, acts on the surface spins. As we have shown, the boundary conditions to be fulfilled by the precession of the surface spins can be expressed by the surface pinning parameter, defined:
\begin{equation}
A=1- \frac{a^2}{D_{ex}}\vec{K}_{surf} \cdot \vec{m},
\label{def_parameter}
\end{equation}
where $a$ is the lattice constant, $D_{ex}$ is the exchange stiffness constant, and $\vec{m}$ denotes a unit vector oriented along the magnetization $\vec{M}$ of the sample. Note that a complete lack of anisotropy field on the surface corresponds to the surface parameter value one; the freedom of the surface spins in this situation will be referred to as the \emph{natural freedom}. In the case of nonzero anisotropy field three situations, substantially different from the physical point of view, may occur depending on the angle between the magnetization~$\vec{M}$ and the surface anisotropy field~$\vec{K}_{surf}$. If the surface spins are aligned perpendicularly to~$\vec{K}_{surf}$, their freedom remains \emph{natural} ($A=1$); otherwise, the surface spins are pinned (and $A<1$) or unpinned (and $A>1$) for the above-mentioned angle acute or obtuse, respectively. All three pinning regimes are schematically depicted in Fig.~\ref{pinning_regimes}.

In the rigorous theory of SWR the surface pinning parameter can be represented (Cracknell and Puszkarski \cite{HP1977,crack}) 
as a series expansion in spherical harmonics $Y_{lm}(\theta,\phi)$:
\begin{align}
\nonumber A\left(\theta,\phi\right) =
1- \frac{a^2}{D_{ex}}\vec{K}_{surf} (\theta, \phi ) \cdot \vec{m} \\ 
\nonumber 
=\sum_{l=0}^\infty \sum_{m=-l}^l A_{lm}Y_{lm}\left(\theta,\phi\right) \\
\nonumber 
=\sum_{l=0}^\infty \left[ a_l P_l^0 \left(\cos{\theta}\right)+ 
 \sum_{m=-l}^l P_l^m\left(\cos{\theta}\right)\right. \\
\left.\times (\alpha_{lm}\cos{m\phi}+\beta_{lm}\sin{m\phi})\right],
\label{harmonics}
\end{align}
where $\theta$ and $\phi$  are the out-of-plane polar angle and the in-plane azimuth angle, respectively, of the magnetization $\vec{M}$. The coefficients $a_l$, $\alpha_{lm}$ and $\beta_{lm}$ (which can be found experimentally) determine the respective energy contributions brought to the effective surface pinning by different surface interactions. As established in Ref.~\onlinecite{crack}, in the case of surface cut (100) -- which is that of the thin-film samples considered in Ref.~\onlinecite{liu2007} -- all the terms with odd values of $l$ vanish, and the only values allowed to $m$ are 0, 4, 8, \ldots{}. In our research we have also observed \cite{HP1979} 
that in the case of thin films the series (\ref{harmonics}) can be cut to only include terms up to $l = 4$, since further contributions tend to be minor. Thus, we propose the following angular dependence of the surface parameter as appropriate for the interpretation of the SWR spectra obtained in Ref.~\onlinecite{liu2007}: 
\begin{eqnarray}
\nonumber A\left(\theta,\phi\right) = 1-a_0-a_2\left(\theta,\phi\right)\left(3\cos^2{\theta}-1\right)\\
-a_4\left(\theta,\phi\right)\cos4{\phi}.
\label{2_3}
\end{eqnarray}
The above formula provides the basis for the elucidation of the most important surface mechanisms behind the SWR surface dynamics in (Ga,Mn)As thin films, which is the main goal of the present paper.

\section{Out-of-plane angle dependence of the surface parameter in (G\lowercase{a},M\lowercase{n})A\lowercase{s} thin films}\label{outofplane}

In the present paper we shall focus on the configuration dependence of the SWR spectrum of (Ga,Mn)As thin films with the external field $\vec{H}$ only rotating in a plane perpendicular to the surface of the sample from the direction along the surface normal ($\theta_{H}=0$) to the in-plane direction ($\theta_{H}=90^\circ$). According to the formula~(\ref{2_3}), in this case the surface parameter of a (Ga,Mn)As thin film can be represented as the series:
\begin{equation}
A\left(\theta_{M}\right)=1-a_{0} -a_{2}\left(\theta_{M}\right)\left(3\cos^2{\theta_{M}}-1\right),
\label{series}
\end{equation}
where $\theta_{M}$ is the angle between the surface normal and the magnetization $\vec{M}$ of the film (let us remark in advance that, except for two extreme configurations, in general $\theta_{H} \neq \theta_{M}$; the relation between $\theta_{H}$ and $\theta_{M}$ will be discussed in detail in the next Section). Note that the adoption of the formula (\ref{series}) implies taking into account only two mechanisms of surface spin pinning: the \emph{isotropic} pinning component $a_{0}$, the influence of which on the freedom of the spins is independent of their configuration with respect to the surface of the film, and the \emph{uniaxial} factor $a_{2}(\theta_{M})$ representing the contribution of the uniaxial symmetry, with the surface normal as the symmetry axis, to the surface pinning. 

Already at this stage interesting conclusions regarding the properties of the surface pinning can be drawn from the equation (\ref{series}) despite its rather general formulation. Let us define two special angles: the \emph{critical angle} $\theta_{M}^{c}$, for which \emph{natural} pinning conditions prevail on the surface of the film, i.e.~$A(\theta_{M}^{c}) \equiv 1$, and the uniaxial pinning \emph{annihilation angle} $\theta_{M}^{u}$, for which the uniaxial pinning vanishes, i.e.~$3\cos^2{\theta_{M}}-1 \equiv 0$. The following equations apply to these special angles:
\begin{subequations}
\begin{equation}
A(\theta_{M}^c) \equiv 1,
\label{critical}
\end{equation}
\begin{equation}
A(\theta_{M}^u) \equiv 1-a_{0}.
\label{uniaxial}
\end{equation}
\end{subequations}
The latter equation provides a simple formula for the determination of the isotropic component $a_{0}$ of the surface pinning, only necessitating the value of the surface parameter in the external field configuration corresponding to the uniaxial pinning annihilation angle~$\theta_{M}^{u}$. With~$a_{0}$ known, the configuration dependence of the uniaxial factor~$a_{2}(\theta_{M})$ can be determined by the measurement of the surface parameter~$A(\theta_{M})$ vs.~$ \theta_{M}$ (see the equation~(\ref{series})). (We shall refer in this regard to the paper by Liu~\textit{et~al.} \cite{liu2007} providing measurement data which will allow us to plot the experimental~$A(\theta_{M})$ dependence; see Section~\ref{model} below.) On the other hand, theoretical considerations within the model used for describing the surface anisotropy in (Ga,Mn)As samples will lead us to an equation, formulated in the next Section, in which~$a_{2}(\theta_{M})$ is expressed by magnetic characteristics of the (Ga,Mn)As thin film. In Section~\ref{confrontation} very interesting conclusions regarding the interrelation between the ranges of the exchange interaction on the surface and in the bulk of (Ga,Mn)As thin films will be drawn from the confrontation of the theory with the experiment. 

\section{Model of the uniaxial surface anisotropy in (G\lowercase{a},M\lowercase{n})A\lowercase{s} thin films}\label{model}

We shall derive a phenomenological formula for the coefficient $a_{2}$ on the basis of our calculations presented in Appendix~\ref{B}, in which the model of the uniaxial anisotropy is considered in both the microscopic and macroscopic approaches. From the equation (\ref{last}) in Appendix~\ref{B} (see also Ref.~\onlinecite{HP+MKasp}) it follows that the coefficient~$a_{2}(\theta_{M})$ in the equation~(\ref{series}) can be expressed~as:
\begin{equation}
a_{2}\left(\theta_{M}\right) = \frac{1}2\left[4\pi\left(M_{eff}^{bulk}-M_{eff}^{surface}\right)\frac{a^{2}}D_{ex}\right],
\label{a2}
\end{equation}
where $4\pi M_{eff} \equiv 4\pi M - H_{2 \bot}$,~$M$ is the saturation magnetization,~$H_{2 \bot}$ the effective uniaxial anisotropy field, $a$~the lattice constant (the average Mn-Mn distance), and~$D_{ex}$ the exchange stiffness constant.
The above equation indicates that both the intrinsic uniaxial anisotropy and the demagnetizing field contribute to the total uniaxial anisotropy in our model. 

As we will see later, extremely informative for the physical interpretation of the experiments performed by Liu~{\textit{et~al.} \cite{liu2007} is the expression of the latter contribution by the exchange length~$\lambda$, defined:
\begin{equation}
\lambda_{b} \equiv \sqrt{\frac{D_{ex}}{4\pi M^{bulk}}}, \text{ }\lambda_{s} \equiv \sqrt{\frac{D_{ex}}{4\pi M^{surface}}};
\label{lambda}
\end{equation}
we have introduced here a \emph{locally} defined exchange length, different for the bulk and the surface. From the physical point of view it is reasonable to assume here that the lattice constant $a$ in the equation (\ref{a2}) is identical with the exchange length $\lambda_{b}$ that characterizes the interaction in the \emph{whole} sample except for its surface. Under these assumptions~(\ref{a2}) becomes: 
\begin{subequations}
\begin{equation}
a_{2}\left(\theta_{M}\right) = a_{2}^{0}+a_{2}^{1} \left(\theta_{M}\right),
\label{a2bis}
\end{equation}
\begin{equation}
a_{2}^{0} \equiv \frac{1}2 \frac{\lambda_{b}^2}{D_{ex}} \left(H_{2 \bot}^{surface} - H_{2 \bot}^{bulk} \right),
\label{a2^0}
\end{equation}
\begin{equation}
a_{2}^{1}\left(\theta_{M}\right) = \frac{1}2 \left[1-\Big(\frac {\lambda_{b}}{\lambda_{s}}\Big)^{2}\right].
\label{a2^1}
\end{equation}
\end{subequations}
In the equations (\ref{a2bis})--(\ref{a2^1}) we have indicated in advance what will follow from the confrontation of these formulas with the experimental data: that only the term~$a_{2}^{1}(\theta_{M})$ is \emph{configuration-dependent}!

\section{Confrontation of our surface pinning model with the SWR study by Liu~\textit{et~al.} \cite{liu2007}}\label{confrontation}

Finally, the formula for the surface parameter takes the form:
\begin{equation}
A\left(\theta_{M}\right)=1-a_{0} - \left[a_{2}^0 + a_{2}^1 \left(\theta_{M}\right)\right]\left(3\cos^2{\theta_{M}}-1\right),
\label{surf_parameter}
\end{equation}
where the coefficients $a_{2}^0$ and $a_{2}^1 \left(\theta_{M}\right)$ are as defined in~(\ref{a2^0}) and~(\ref{a2^1}). Note that in the surface inhomogeneity model the surface parameter~(\ref{surf_parameter}) measures the degree of pinning of the surface spins and describes quantitatively the degree of the dynamic freedom with which they participate in the motion of the whole system of spins. The value~$A=1$ corresponds to a special case referred to as the \emph{natural} freedom of the surface spins. Acquired by the surface spins as a result of breaking their interaction with those of their neighbors which are eliminated by the introduction of the surface, this freedom stems \emph{solely} from the broken symmetry in the vicinity of the surface spins. Thus, \emph{absolute natural freedom} of the surface spins only occurs when all the energy contributions in the equation~(\ref{surf_parameter}) vanish \emph{simultaneously},~i.e.:  
\begin{subequations}
\begin{equation}
a_0 \equiv 0,
\end{equation}
\begin{equation}
H_{2 \bot}^{surface} = H_{2 \bot}^{bulk},
\end{equation}
\begin{equation}
\lambda_{b}=\lambda_{s}.
\label{abs_freedom_c}
\end{equation}
\label{abs_freedom}
\end{subequations}
However, as confirmed experimentally, the natural freedom of the surface spins is possible also in a \emph{particular situation} in which the surface parameter value is one even though the conditions (\ref{abs_freedom})
are not all fulfilled. This particular situation may occur when there exists such a \emph{critical angle} $\theta_{M}^c$ that $A\left(\theta_{M}^c\right)\equiv 1$ because all the energy contributions in (\ref{surf_parameter}) annihilate each other. Further in this Section we shall analyze this situation in detail.

On the basis of their SWR study of (Ga,Mn)As thin films Liu~\textit{et~al.}\cite{liu2007} plotted the configuration dependence of the surface parameter~$A\left(\theta_{H}\right)$ with the \emph{magnetic field} rotating from the perpendicular ($\theta_{H}=0$) to parallel ($\theta_{H}=90^\circ$) configuration (see Fig.~9 in Ref.~\onlinecite{liu2007}). As our formula (\ref{surf_parameter}) concerns the configuration dependence of the surface parameter versus~$\theta_{M}$, i.e. with rotating \emph{magnetization} of the sample, the first thing necessary for proper interpretation of the measurements of Liu~\textit{et~al.} was to find the dependence~$\theta_{M}=\theta_{M} \left(\theta_{H}\right)$ in equilibrium conditions. The determination of the equilibrium conditions and the derivation of the sought relation~$\theta_{M}=\theta_{M} \left(\theta_{H}\right)$ between the two configuration angles are presented in Appendix~\ref{A}. Figure~\ref{parameter} shows the recalculated configuration dependence of the surface parameter, with~$A$ plotted versus the new variable~$\theta_{M}$; the plot corresponds to the measurement data of Liu~\textit{et~al.} presented in Ref.~\onlinecite{liu2007}, Fig.~9. The natural surface pinning is seen to occur for the critical angle~$\theta_{M}=35^\circ$ (which corresponds to the experimental angle $\theta_{H}=19^\circ$). Also, the new plot reveals the occurrence of a local maximum in the $A\left(\theta_{M}\right)$ dependence around the angle $\theta_{M}^u=54.73^\circ$, for which the term $\left(3\cos^2{\theta_{M}^u}-1\right)$ equals zero (we shall take advantage of this finding below in further analysis of the experimental data of Liu~\textit{et~al.}.\cite{liu2007})

\begin{figure}
\includegraphics[width=6 cm,  angle=270]{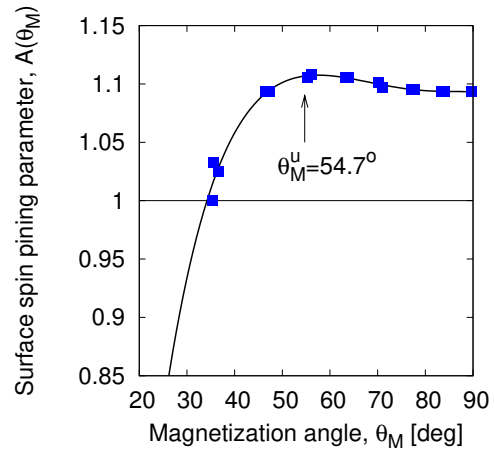}
\caption{Magnetization angle dependence of the surface pinning parameter~$A\left(\theta_M\right)$ according to the experimental data obtained by Liu~\textit{et~al.} \cite{liu2007} in their SWR study of a (Ga,Mn)As thin film; the plot corresponds to that shown in Fig.~9a in the cited paper, presenting the dependence on the magnetic field angle~$\theta_H$. The  applied transformation between the angles~$\theta_H$ and ~$\theta_M$ is based on our determination of the equilibrium direction of the magnetization, presented in Appendix~A.}
\label{parameter}
\end{figure}

Now we will demonstrate that the experimental curve shown in Fig.~\ref{parameter} 
can be described by the function resulting from our SI model:
\begin{equation}
A\left (\theta_{M}\right)=1-a_{0} -a_{2}\left(\theta_{M}\right) \left (3\cos^2{\theta_{M}}-1\right).
\label{series_SI}
\end{equation}
Knowing the maximal value of the surface parameter, $A(\theta_{M}^u)=1.1068$, we obtain immediately the value of the isotropic term in the series (\ref{series_SI}):
\begin{equation}
a_{0}=-0.1068.
\label{a0_value}
\end{equation} 
On the other hand, the condition of occurrence of the local maximum at $\theta_{M}^u$ implies that the coefficient $ a_{2}\left(\theta_{M}\right)$ is zero at this point: 
\begin{equation}
a_{2}\left(\theta_{M}^u\right)=0.
\end{equation}
Both conditions allow to determine \emph{explicitly} the function $a_{2}\left(\theta_{M}\right)$ that reproduces the experimental plot shown in Fig.~\ref{parameter}
 via the series (\ref{series_SI}). The determined function $a_{2}\left(\theta_{M}\right)$  is presented in Fig.~\ref{coefficient}.

\begin{figure}
\includegraphics[width=6 cm, angle=270]{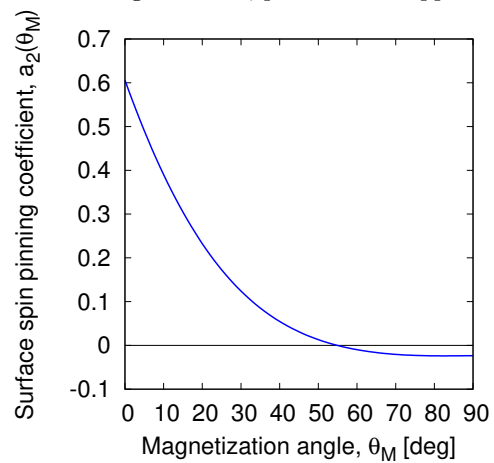}
\caption{Magnetization angle dependence of the surface pinning coefficient $a_2\left(\theta_M\right)$ calculated from Eq.~(\ref{series_SI}).}
\label{coefficient}
\end{figure}

In the next step we shall refer to the formula~(\ref{a2bis}) postulated in our model and representing the coefficient~$a_{2}\left(\theta_{M}\right)$ as the sum of a \emph{constant} component~$a_{2}^0$ and a \emph{function}~$a_{2}^1\left(\theta_{M}\right)$. This implies that $a_{2}\left(\theta_{M}\right)$ and $a_{2}^1\left(\theta_{M}\right)$ have the same angular dependence, and their plots only differ by a shift $a_{2}^0$ along the ordinate axis. However, we do not know the value of~$a_{2}^0$! This is a very sensitive point of our considerations, since in order to establish the value of $a_{2}^0$ we have to refer to the physical assumptions that are the very basis of our model of surface anisotropy. It seems reasonable to assume that of the three conditions (\ref{abs_freedom}) only (\ref{abs_freedom_c}) is fulfilled in the critical angle configuration; the other two energy contributions do not vanish, but compensate each other. This assumption means that by virtue of the equation~(\ref{a2^1}) the coefficient $a_{2}^1$ vanishes in the critical angle configuration:
\begin{equation}
a_{2}^1\left(\theta_{M}^c\right)\equiv 0,
\end{equation}
which implies the equality:
\begin{equation}
a_{2}^0 = a_{2}\left(\theta_{M}^c\right).
\label{a2^0_value}
\end{equation}
Having established the value of the component $a_{2}^0$ we can already determine \emph{explicitly} the function $a_{2}^1\left(\theta_{M}\right)$. The result is shown in Fig.~\ref{coefficient1}. 

\begin{figure}
\includegraphics[width=6 cm, angle=270]{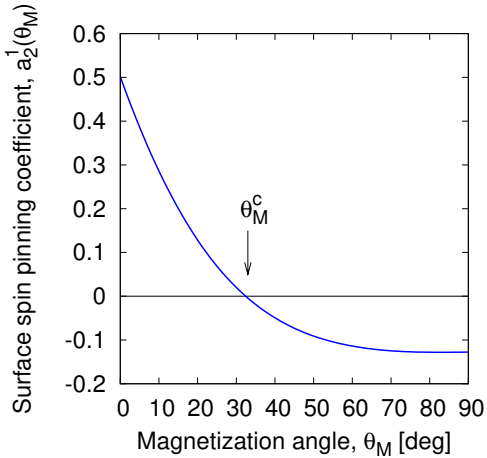}
\caption{Magnetization angle dependence of the surface pinning coefficient $a_2^1\left(\theta_M\right)$ calculated from Eq.~(\ref{surf_parameter}). (See the text for detailed discussion.)}
\label{coefficient1}
\end{figure}

From (\ref{abs_freedom_c}) it follows that:
\begin{equation}
\frac{\lambda_{s}}{\lambda_{b}}=\frac{1}{\sqrt{1-2a_2^1\left(\theta_M\right)}},
\label{lambda_ratio}
\end{equation}
and, on the basis of Fig.~\ref{coefficient1}, we can find the~$\theta_M$ dependence of the $\lambda_{s}/\lambda_{b}$ ratio. The obtained dependence is shown in Fig.~\ref{ratio}. Its analysis leads to very interesting physical conclusions.

\begin{figure}
\includegraphics[width=6 cm, angle=270]{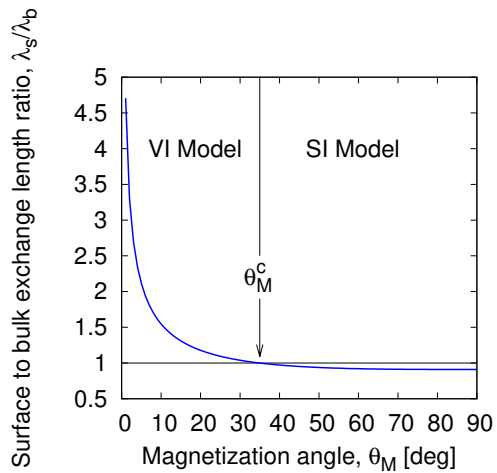}
\caption{Magnetization angle dependence of the~$\lambda_s/\lambda_b$ ratio resulting from our model of surface pinning in (Ga,Mn)As thin films (see Eq.~(\ref{lambda_ratio})); $\lambda_s$ and $\lambda_b$ denote the surface and bulk exchange length, respectively;~$\theta_M^c$ is the SWR critical angle.}
\label{ratio}
\end{figure}

Note that in the plot in Fig.~\ref{ratio} the surface exchange length $\lambda_{s}$ is only slightly smaller than the bulk exchange length $\lambda_{b}$ for any angle $\theta_{M}$ between the critical angle $\theta_{M}^c$ and the parallel configuration angle $\theta_{M}=90^\circ$:
\begin{equation}
\theta_{M}^c < \theta_{M}\leq 90^\circ.
\label{angle}
\end{equation}

Thus, in this angle range a surface disturbance will not go beyond the first sub-surface plane formed by the spins directly under the surface. This means that the assumptions of the SI model are fulfilled very well in the angle range defined by (\ref{angle})! In contrast, for angles $\theta_{M} < \theta_{M}^c$ $\lambda_{s}$ is greater than $\lambda_{b}$ and grows steeply as the perpendicular configuration $\theta_{M}=0$ is approached. This means that in this angle range a surface disturbance, rather than being localized at the surface, penetrates into the bulk, affecting deeper sub-surface planes. Thus, the applicability of the SI model is very limited in this angle range, and the volume inhomogeneity model will be more adequate. This conclusion is fully confirmed by the experimental study by Liu~\textit{et~al.}\cite{liu2007}

\section{Further physical implications of the model}\label{further_implications}

Now let us consider the component $a_{2}^0$, which we have found to have a constant value, specified in (\ref{a2^0_value}), 
throughout the angle range $\theta_{M}\in \left(0,\pi /2\right) $. From the derived formula~(\ref{a2^0}) for~$a_{2}^0$ it follows that its constant value implies the invariance of $\lambda_b$ in the rotation of the magnetization of the sample (we have already taken advantage of this fact, interpreting the angular dependence of the~$\lambda_s / \lambda_b$ ratio as only due to~$\lambda_s$ in the preceding Section).  The~measurements performed by Liu~\textit{et~al.} indicate that the material parameter values in the studied (Ga,Mn)As sample are~$D_{ex}=3.79 \text{ T}\cdot \text{nm}^2$ and~$4\pi M_{eff}=4588\text{ Oe}$, implying~$\lambda_b \approx 3 \text{ nm}$. On the other hand, for the critical angle~$\theta_{M}=\theta_{M}^c$ from the formula (\ref{surf_parameter}) we get the equality:
\begin{equation}
a_0+a_2^0 \left(3\cos^2{\theta_{M}^c}-1\right)=0,
\end{equation}
which, after the substitution of $a_{0}=-0.1068$ and $\theta_{M}^c =$~$35^\circ$,
yields the sought value:
\begin{equation}
a_2^0 \cong 0.108.
\end{equation}
Now, getting back to (\ref{a2^0}), with the above-determined value of~$a_2^0$ we can estimate the difference between the effective uniaxial anisotropy field values on the surface and in the bulk:
\begin{equation}
\Delta H_{2 \bot} \equiv H_{2 \bot}^{surface} - H_{2 \bot}^{bulk} \approx 913 \text{ Oe}.
\end{equation}
To our best knowledge, this is the first quantitative estimate of the surface uniaxial anisotropy field in (Ga,Mn)As thin films to be reported in the literature.

As a measure of surface spin pinning experimentalists tend to use the surface anisotropy energy $E_s\left(\theta_M\right)$, a phenomenological quantity thus related to the surface pinning parameter $A$ used by us for describing the same feature:
\begin{equation}
E_s\left(\theta_M\right)=\frac{M D_{ex}}{\lambda_b}\left[A\left(\theta_M\right)-1\right].
\label{anis_const}
\end{equation}
The above relation indicates that the character of the angular dependence of both quantities used for describing the surface pinning is identical, though in the equation~(\ref{anis_const}) the reference level of the measure of the surface pinning is
the zero value of the surface anisotropy energy, corresponding to our natural pinning 
$A=1$.
For~$E_s\left(\theta_M\right)>0$ the surface spins are \emph{unpinned}, while for~$E_s\left(\theta_M\right)<0$ their freedom is constrained, which means that the surface spins are \emph{pinned}. Plotted in Fig.~\ref{ani_energy}, 
$E_s\left(\theta_M\right)$ has a maximum for~$\theta_M=\theta_M^u$; according to our estimate its maximal value is~$E_s\left(\theta_M^u\right) \approx$~$0.07 \text{ erg}/ \text{cm}^2$. Note that this maximal surface anisotropy value is solely related to the free component $a_0$, only responsible for the isotropic part of the surface spin pinning; the other surface anisotropy components only reduce this (maximal) value as the angle diverges from~$\theta_M^u$ in either direction.

\begin{figure}
\includegraphics[width=6 cm, angle=270]{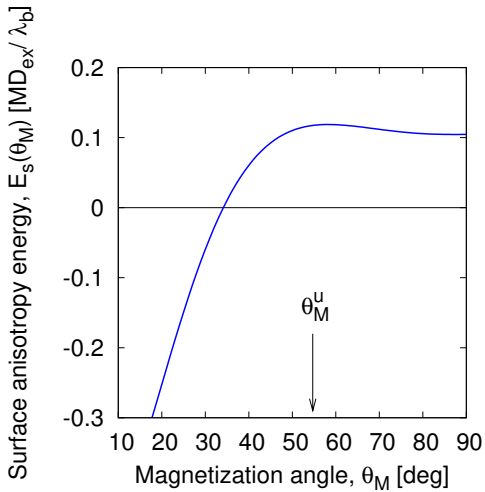}
\caption{Magnetization angle dependence of the surface anisotropy energy $E_s\left(\theta_M\right)$ resulting from our theory (see Eq.~(\ref{anis_const})) for the (Ga,Mn)As thin film investigated by Liu~\textit{et~al.}\cite{liu2007}}
\label{ani_energy}
\end{figure}

\section{Final remarks}\label{final}
In our model the SWR critical angle is determined from the condition that the exchange length must be the same on the surface and in the bulk:
\begin{equation} 
\lambda_s = \lambda_{b}.
\end{equation}

On the other hand, the experimental studies indicate that $\lambda_{b}$ is configuration-independent, and only the surface exchange length $\lambda_{s}$ is sensitive to the configuration of the magnetization of the film with respect to its surface. We suggest that this might be related to the fact that also the charge carrier (hole) concentration on the surface is different than in the bulk in the studied material. \cite{ohms, nishi} If this hypothesis of ours is true, then any experimental treatment modifying the charge carrier concentration on the \emph{surface} of the studied sample should alter the SWR critical angle! This may be performed for instance by \emph{hydrogenation} of the sample, since short-time hydrogenation has been shown \cite{bihler2009} to provide an efficient tool for manipulating the effective surface spin pinning by changing the hole concentration profile of the sample. The suggested experiment would provide a direct proof that the range of the exchange interaction in the ferromagnetic semiconductor (Ga,Mn)As is correlated with the charge carrier concentration.

\section{Summary}

In this  paper we show
 why in different 
conditions a resonance (Ga,Mn)As thin film sample may meet the assumptions of either the Surface Inhomogeneity (SI)
or the Volume Inhomogeneity (VI) model. 
In our considerations we refer 
to the spin-wave resonance (SWR) spectra measured by X.~Liu~\textit{et~al.} \cite{liu2007}
in (Ga,Mn)As thin films in different configurations of the static magnetic field $\vec{H}$ with respect to the surface. 
We demonstrated that the observed configuration dependence of the SWR spectrum of the studied material can be described 
with the use of the surface pinning parameter expressed by the formula:
\begin{equation}
\nonumber
A\left(\theta_{M}\right)=1-a_{0} - \left[a_{2}^0 + a_{2}^1 \left(\theta_{M}\right)\right]\left(3\cos^2{\theta_{M}}-1\right),
\label{surf_parameter}
\end{equation}
where $\theta_{M}$ is the angle between the surface normal and the magnetization $\vec{M}$ of the sample. The values 
of the coefficients are estimated on the basis of the experimental data; the estimated value of the \emph{isotropic} 
component of the surface pinning, $a_0 = -0.1068$, allows to determine the maximal surface anisotropy energy density, 
$E_s \approx 0.07 \text{ erg}/\text{cm}^2$. The intrinsic \emph{uniaxial} anisotropy term $a_2^0$ is of the order of~0.1, 
which implies that the uniaxial anisotropy field~$H_{2\bot}$ on the surface exceeds the bulk value by ca.~0.1~T. 
We postulated that the coefficient $a_{2}^1 \left(\theta_{M}\right)$ is related to the difference between the surface 
and bulk exchange lengths ($\lambda_s$ and $\lambda_b$, respectively), which, when confronted with the measurements, 
 implies that (unlike $\lambda_{b}$) only $\lambda_{s}$ depends on $\theta_M$, 
or the magnetization configuration with respect to the surface. For a critical angle $\theta_M^c$, at which the 
SWR spectrum collapses to a \emph{single} peak, $\lambda_s = \lambda_b$. For angles $\theta_M > \theta_M^c$ the surface 
exchange length $\lambda_s$ is slightly smaller than the bulk exchange length $\lambda_b$: $\lambda_s < \lambda_b$, 
whereas for $\theta_M < \theta_M^c$ 
$\lambda_s$ is greater than~$\lambda_b$ and grows steeply as the perpendicular configuration ($\theta_M = 0$) is approached. 
This finding shows that the critical angle $\theta_M^c$ separates two angle ranges in which the resonance properties 
are different: for $\theta_M > \theta_M^c$ the SI model applies, since $\lambda_s \approx \lambda_b$, and 
for $\theta_M < \theta_M^c$ the VI model is adequate due to the domination of the surface exchange length ($\lambda_s \gg \lambda_b$). 
Seeking the physical grounds of this result, we proposed a working hypothesis that the discovered property is correlated 
with inhomogeneous distribution of the concentration of holes mediating the long-range magnetic interaction between 
localized spins along the surface normal. We suggested further experiments to verify this hypothesis.

\begin{acknowledgments}
This study is a part of a project financed by \mbox{Narodowe} Centrum Nauki (National Science Centre of Poland), Grant no. DEC-2013/08/M/ST3/00967.
\mbox{Henryk}~Puszkarski would like to address special thanks to Prof.~J.~K.~Furdyna of Notre Dame University for the support given to the project and his kind interest in this work.
The authors are also much indebted to Professor A.R. Ferchmin for highly usefull discussions and for 
critical reading the manuscript.
\end{acknowledgments}


\appendix
\section{Determination of the equilibrium direction of magnetization in (Ga,Mn)As thin films}\label{A}

The experimental SWR spectra analyzed in this paper were measured in the \textquotedblleft out-of-plane geometry\textquotedblright, as referred to by the Authors of Ref.~\onlinecite{liu2007}. In this out-of-plane geometry, the (Ga,Mn)As layer was cemented to a parallelepiped of GaAs (100) substrate material, the [110] edge of the specimen oriented vertically. The external magnetic field~$\vec{H}$ was confined to the horizontal plane (i.e. perpendicular to the film surface) allowing SWR measurements with $\vec{H}$ in any intermediate orientation between the normal to the film surface, $\vec{H}\parallel[001]$,
and the in-plane orientation, $\vec{H}\parallel[1\bar{1}0]$.
In this particular geometry the magnetization $\vec{M}$ of the sample lies in the same horizontal plane as the field $\vec{H}$. Thus, the spatial orientation of the vectors $\vec{H}$ and $\vec{M}$ is defined by two polar angles, $\theta_{H}$ and $\theta_{M}$, between the respective vectors and the normal to the surface of the film. For (Ga,Mn)As samples in this particular geometry of the external field the free energy density $F_{\bot}$ of the system has the form:\cite{zhou2009}
\begin{multline}
F_{\bot}=\frac{1}{2} M \times \bigg[-2H\left(\cos{\theta_M}\cos{\theta_H}+\sin{\theta_M}\sin{\theta_H}\right) \\
\left.+\left(4\pi M-H_{2\bot}\right)\cos^2{\theta_M}-\frac{1}{2}H_{4\bot}\cos^4{\theta_M}\right.\\
\left.-\frac{1}{4}H_{4\parallel}\sin^4{\theta_M}-H_{2\parallel}\sin^2{\theta_M}\right],
\label{free}
\end{multline}
where $H_{2\bot}$ and $H_{4\bot}$ are the uniaxial and cubic anisotropy fields, respectively, perpendicular to the plane of the sample; $H_{2\parallel}$ and $H_{4\parallel}$ are the in-plane uniaxial and cubic anisotropy fields, respectively.
In the investigated (Ga,Mn)As sample these four bulk parameters have the values:\cite{liu2007}
\begin{subequations}
\begin{equation}
4\pi M_{eff} \equiv 4\pi M-H_{2\bot}=4588 \text{ Oe},
\end{equation}
\begin{equation}
H_{4\bot}=0, H_{4\parallel}=197\text{ Oe}, H_{2\parallel}=77\text{ Oe}.
\end{equation}
\end{subequations}
Let us determine now the equilibrium direction of the magnetization of the sample, i.e. the equilibrium angle~$\theta_M$. We will use the condition of equilibrium of the system, which requires the first derivative of its free energy~$F_{\bot}$ to vanish:
\begin{equation}
\frac{\partial{F_{\bot}}}{\partial{\theta_M}}=0;
\label{equilib}
\end{equation}
this condition allows to determine the sought relation~$\theta_M=\theta_M\left(\theta_H\right)$.
\begin{figure}[h]
	\centering
		\includegraphics[width=6 cm, angle=270]{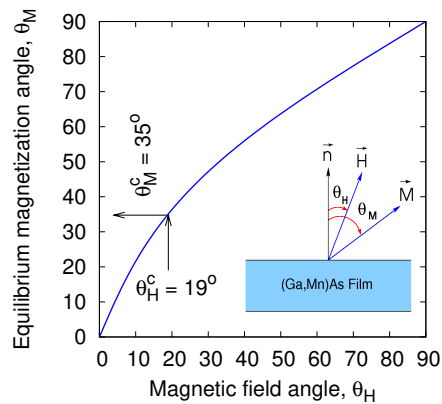}
	\caption{Equilibrium magnetization angle $\theta_M$ vs. the external field angle $\theta_H$ as determined from the condition~(\ref{equilib}) 
for the (Ga,Mn)As thin film studied by Liu~\textit{et~al.};\cite{liu2007} $\theta_M^c$ and $\theta_H^c$ are the respective critical SWR angles.}
	\label{equilibrium}
\end{figure}

Since the condition (\ref{equilib}) must be fulfilled when resonance occurs, the magnetic field~$H$ in (\ref{free}) is the resonance field, $H\equiv H_{res}$; we read its value from Fig.~5 in Ref.~\onlinecite{liu2007}, identifying it with the resonance field of the \emph{fundamental} mode ($n=1$). The $\theta_M=\theta_M\left(\theta_H\right)$ relation determined numerically on the basis of the above considerations is shown in Fig.~\ref{equilibrium}; we refer to this relation many times in this paper when analyzing the experimental SWR spectra reported by Liu~\textit{et~al.} \cite{liu2007}

\section{Surface vs. bulk uniaxial anisotropy}\label{B}

In this Appendix we shall consider the case in which only the perpendicular uniaxial anisotropy $H_{2\bot}$ enters the formula (\ref{free}) for the free energy. In that case the free energy reads:
\begin{multline}
F_{\bot}=\frac{1}{2} M \times \bigg[-2H\left(\cos{\theta_M}\cos{\theta_H}+\sin{\theta_M}\sin{\theta_H}\right) \\
+4\pi M_{eff}\cos^2{\theta_M}\bigg],
\label{free_perp}
\end{multline}
and the use of the well-known Smit-Beljers resonance formula:
\begin{multline}
\left(\frac{\omega}{\gamma}\right)^2 = \frac{1}{M^2\sin^2{\theta_M}}
\Bigg[\frac{\partial^2{F_{\bot}}}{\partial{\phi_M^2}}
\frac{\partial^2{F_{\bot}}}{\partial{\theta_M^2}}\Bigg. \\
\left.-\left(\frac{\partial^2{F_{\bot}}}{\partial{\phi_M}\partial{\theta_M}}\right)^2\right]
\label{smit}
\end{multline}
leads to the following configuration resonance condition, only applying to the uniform mode $k_{\bot}\equiv 0$ in the case considered:  
\begin{multline}
\left(\frac{\omega}{\gamma}\right)^2=\left[H\cos{\left(\theta_M-\theta_H\right)}-4\pi M_{eff}\cos{2\theta_M}\right] \\
\times \left[H\cos{\left(\theta_M-\theta_H\right)}-4\pi M_{eff}\cos^2{\theta_M}\right].
\label{res_cond}
\end{multline}
It will be very informative to derive the same condition in the microscopic approach, in which the energy of the system is expressed by the Hamiltonian:
\begin{multline}
\hat{\mathscr{H}}=-J\sum_{l\vec{j}\neq l'\vec{j}'}\hat{S}_{l\vec{j}}\cdot\hat{S}_{l'\vec{j}'}-g\mu_B\sum_{l\vec{j}}
\vec{H}\cdot \hat{S}_{l\vec{j}} \\
-D\sum_{l\vec{j}}\left(\hat{S}^z_{l\vec{j}}\right)^2;
\label{hamil}
\end{multline}
its successive terms account for the isotropic exchange interaction, the Zeeman energy of the spins, and the perpendicular uniaxial anisotropy energy. The subscript $l\vec{j}$ defines the position of the given spin, with $l$ labeling the layer and the two-dimensional vector $\vec{j}$ defining the position of the spin $\hat{S}_{l\vec{j}}$ in the $l$-th layer. The energy of a standing spin wave with a wave number $k_{\bot}$ in this model is given by the expression: \cite{HP79}
\begin{align}
\nonumber \left(\frac{\omega}{\gamma}\right)^2=\left[H\cos{\left(\theta_M-\theta_H\right)}+\frac{2DS}{g\mu_B}\cos{2\theta_M}\right. \\
\nonumber \left.+\frac{2Sz_{\bot}Ja^2}{g\mu_B}k_{\bot}^2\right] \\
\nonumber \times \left[H\cos{\left(\theta_M-\theta_H\right)}+\frac{2DS}{g\mu_B}\cos^2{\theta_M}\right. \\
\left.+\frac{2Sz_{\bot}Ja^2}{g\mu_B}k_{\bot}^2\right].
\label{res_cond_micro}
\end{align}
This condition is the counterpart of the condition (\ref{res_cond}) obtained in the macroscopic approach (for $k_{\bot}\neq 0$). From the comparison of these two formulas it follows that:
\begin{subequations}
\begin{equation}
4\pi M_{eff}\equiv -\frac{2DS}{g\mu_B}
\end{equation}
and the coefficient at $k_{\bot}^2$ can be identified as:
\begin{equation}
D_{ex}\equiv \frac{2Sz_{\bot}Ja^2}{g\mu_B}.
\end{equation}
\label{ident}
\end{subequations}

To obtain the formula for the surface parameter expressed by the surface perpendicular uniaxial anisotropy, we must yet rewrite the third term in the Hamiltonian~(\ref{hamil}) in the generalized form:
\begin{equation}
\hat{\mathscr{H}}_a=- \sum_{l\vec{j}}D_{l}\left(\hat{S}^z_{l\vec{j}}\right)^2,
\label{hamil_general}
\end{equation}
where the uniaxial anisotropy constant $D_{l}$ is assumed to be:
\begin{equation}
D_{l}=\left\{
\begin{array}{l}
D_s \text{ for surface spins},\\
D_b \text{ for bulk spins}.
\end{array}
\right.
\end{equation}
On the basis of our earlier papers \cite{HP+MKasp,diep80} it can be demonstrated that in the approximation assuming circular spin precession the following expression for the surface parameter results from this model: 
\begin{equation}
A=1-\frac{D_b-D_s}{2z_{\bot}J}\left(1-3\cos^2{\theta_M}\right),
\end{equation}
where $D_b$ and $D_s$, as indicated above, denote the bulk and surface values, respectively, of the \emph{microscopic} uniaxial anisotropy constant. Now, using the identity relations~(\ref{ident}) we obtain the sought final formula in which the surface parameter is expressed by \textit{macroscopic} quantities:
\begin{align}
\nonumber A\left(\theta_M\right)=1-\frac{1}{2}\left[4\pi\left(M_{eff}^{surface}-M_{eff}^{bulk}\right)\frac{a^2}{D_{ex}}\right] \\
\times \left(1-3\cos^2{\theta_M}\right).
\label{last}
\end{align}



\newpage

\end{document}